# Online Coherence Identification Using Dynamic Time Warping for Controlled Islanding

Hasan Ul Banna, Zhe Yu, Di Shi, Zhiwei Wang, Dawei Su, Chunlei Xu, Sarika Solanki, Jignesh Solanki

***Abstract*—Controlled islanding is considered to be the last countermeasure to prevent system-wide blackouts in case of cascading failures. It splits the system into self-sustained islands to maintain transient stability at the expense of possible loss of load. Generator coherence identification is critical to controlled islanding scheme as it helps identify the optimal cut-set to maintain system transient stability. This paper presents a novel approach for online generator coherency identification using phasor measurement unit (PMU) data and dynamic time warping (DTW). Results from the coherence identification are used to further cluster non-generator buses using spectral clustering with the objective of minimizing power flow disruption. The proposed approach is validated and compared to existing methods on the IEEE 39-bus system, through which its advantages are demonstrated.**

## I. Introduction

With the expansion of power grids in the form of regional interconnections and diverse transmission structure driven by ever-increasing market competition, safe and stable operations of the system have become crucial. Big disturbances, such as natural disasters and human errors, may trigger cascading failures and result in system-wide blackouts, which pose a significant threat to properties and lives [1].

Controlled islanding is an effective approach to prevent system-wide instabilities and blackouts. It splits a power system into smaller subsystems, referred to as islands. The objective is to form stable islands by selecting an optimal set of lines to disconnect while minimizing generation/load imbalance, maintaining voltage stability, ensuring generators coherency, and restraining out-of-step oscillations. The stability of these islands depends on the coherency of generators within each island, which makes a precise and adaptive identification of coherent generators an essential prerequisite. Furthermore, the optimal grouping of generators varies over time, due to changing network topology and operating conditions. Thus the real-time determination of coherency is preferred in practical operations [2]. With the deployment of increasing number of phasor measurement units, online measurement-based coherency identification has become feasible.

There is substantial literature on the coherency identification. Continuation method [3] and eigenvalue analysis approach [4] are applied to different operating conditions. However, both methods require precise knowledge of the system model, which is unavailable in practice. In [5], coherent generator groups are identified using discrete Fourier transform of phasor angle difference of each generator with the center of angle (COA). Internal voltage phasors of generators are estimated by using voltage and current phasors measured by PMUs at generator terminals. Jonsson et al., further improved this method by combining generator speed with Fourier analysis [6]. Inter-area dominant modes are identified as Fourier coefficients with the largest amplitude. However, Fourier analysis based approaches assume linearity and stationarity of the data, which could not be justified when it comes to inter-area oscillations. The principal component analysis (PCA) method proposed in [7] uses bus voltage angle and generator speed for coherency identification. It requires additional prior information of system dynamic characteristics. A correlation coefficient based method was proposed in [8] to overcome the deficiency of PCA method. However, it needs a threshold to identify the correct number of coherent groups, which requires expert knowledge and may vary for different operating conditions and fault locations. Bioinformatics clustering technique is suggested in [9] to determine the coherent groups of generators; however, the number of clusters should be specified which may result in unrealistic grouping if recommended number of clusters is improper. Ariff et al. presented an approach based on independent component analysis and considered 20 sec time window data of generator speeds and bus voltage angles to have reasonable and practical grouping [10]. Another measurement based approach using ANN was presented in [11] which needs excessive offline training to train neurons for online coherent groups identification. For large interconnected networks, consideration of all possible grouping cases is a challenging task of this approach for proper offline training.

On the other hand, the literature on coherency identification with communication loss is less extensive. The explosive number of PMUs poses an enormous burden on communication networks, which may cause link failures. No literature has demonstrated online coherency

This work is funded by SGCC Science and Technology Program under contract no. 5455HJ160007.

H. Banna, Z. Yu, D. Shi, and Z. Wang are with GEIRI North America, 250 W. Tasman Dr., San Jose, CA 95134, United States. (Email: majorhasan_209@yahoo.com, {di.shi, zhe.yu, zhiwei.wang}@geirina.net).

D. Su and C. Xu are with Grid Dispatch Center, State Grid Jiangsu Electric Power Company, Nanjing, China. (Email: 543062082@qq.com, chunleixu@sina.com).

S. Solanki and J. Solanki are with Department of CSEE, West Virginia University, Morgantown, WV, United States. (Email: {Solanki, jignesh.solanki}@mail.wvu.edu).

identification methods that are robust and accurate given delay or partial loss of PMU data.

This paper proposes a coherence identification approach for online implementation which can handle partial observability of the system. It provides an adaptive option to system operators for intentional islanding operation to minimize the impact of cascading outages. Dynamic time warping, which has been extensively used in pattern recognition filed for similarity matching tasks, is employed to cluster generators. The proposed algorithm uses generators rotor angles, estimated through PMU measurements, based on which optimal cut-set can be determined with minimum circuit breaker option and load shedding. The proposed approach has been demonstrated on the IEEE 39-bus system against correlation based [24] and community detection based [2] islanding approaches. Time domain simulations are used to validate and demonstrate the effectiveness of the proposed methodology in minimizing impacts of cascading outages and system-wide blackouts.

The remainder of this paper is organized as follows. Section II presents the problem and describes the proposed generator coherence identification approach. Section III discusses the proposed controlled islanding framework. Simulation results are presented in section IV while conclusions are drawn in section V.

## II. Coherence Identification and Cut-set Determination

When a disturbance occurs in a power system, some generators behave similarly because of their inertia and locations. These generators are considered to be coherent in time domain responses and hence can be clustered in the same group if necessary. The rotor angle response can be selected as the metric for generator coherence identification. Generator $p$ and $q$ are considered coherent if $\Delta\delta_p(t) - \Delta\delta_q(t) \approx 0$ or $\Delta\delta_p(t)-\Delta\delta_q(t)$ = constant, where $\Delta\delta_p(t)$ and $\Delta\delta_q(t)$ are the deviations of rotor angles of generator $p$ and $q$, respectively [12]. In this section, DTW technique is proposed to identify the similarity between rotor angle responses of generators in the system.

### A. DTW Based Generator Coherency Identification

Given voltage and current phasor measurements at $n$ generator terminal buses, rotor angle responses of these generators $\delta$ can be estimated using Least Squares (LS) or Kalman Filter (KF) based approaches [13]. Consider two rotor angle trajectories $\delta_p=\{\delta_{p1}, \delta_{p2}, \delta_{p3},\ldots, \delta_{pi}\}$ and $\delta_q=\{\delta_{q1}, \delta_{q2}, \delta_{q3},\ldots, \delta_{qk}\}$ estimated over the same time period, where $i$ and $k$ are numbers of data points for generators $p$ and $q$, respectively. Normally $i$ and $k$ are equal. When there is data loss or significant communication delays in PMU data transmission, $i$ and $k$ are different, and DTW can still handle the data.

A local distance measure $d(\delta_{pm}, \delta_{qn})$ of points $m$ and $n$ from rotor angle trajectories $\delta_p$ and $\delta_q$ respectively is defined as:

$$d(\delta_{pm},\delta_{qn}) = \left\|\delta_{pm},\delta_{qn}\right\|^2 \tag{1}$$

where, $m \in \{1, 2, 3, \ldots, i\}$ and $n \in \{1, 2, 3, \ldots, k\}$. Similarly, a distance matrix $D(\delta_p, \delta_q)$ of size $i$-by-$k$ is constructed by calculating local distance measures of each pair of data points from trajectories $\delta_p$ and $\delta_p$.

Define $w=\{w_1, w_2, w_3,\ldots, w_L\}$ as a warping path, where $w_l=(m_l, n_l)\in[1{:}i]\times[1{:}k]$ represents the cell in the $m_l$th row, $n_l$th column of a distance matrix $D(\delta_p, \delta_q)$. A valid warping path as shown in Fig. 1 satisfies the following conditions stated in [14]:

- *"**Boundary condition**: $w_1=(1, 1)$ and $w_L=(i, k)$. This condition ensures that the warping path starts and ends at diagonally opposite corner cells of the distance matrix $D(\delta_p, \delta_q)$."*
- *"**Continuity**: if $w_1=(a, b)$ and $w_{l-1}=(a', b')$, $a-a'\leq 1$ and $b-b'\leq 1$. This condition restricts the feasible warping path to be made of only adjacent cells."*
- *"**Monotonicity**: if $w_l=(a, b)$ and $w_{l-1}=(a', b')$, $a-a'\geq 0$ and $b-b'\geq 0$. This condition ensures the path in $w$ to be made monotonically."*

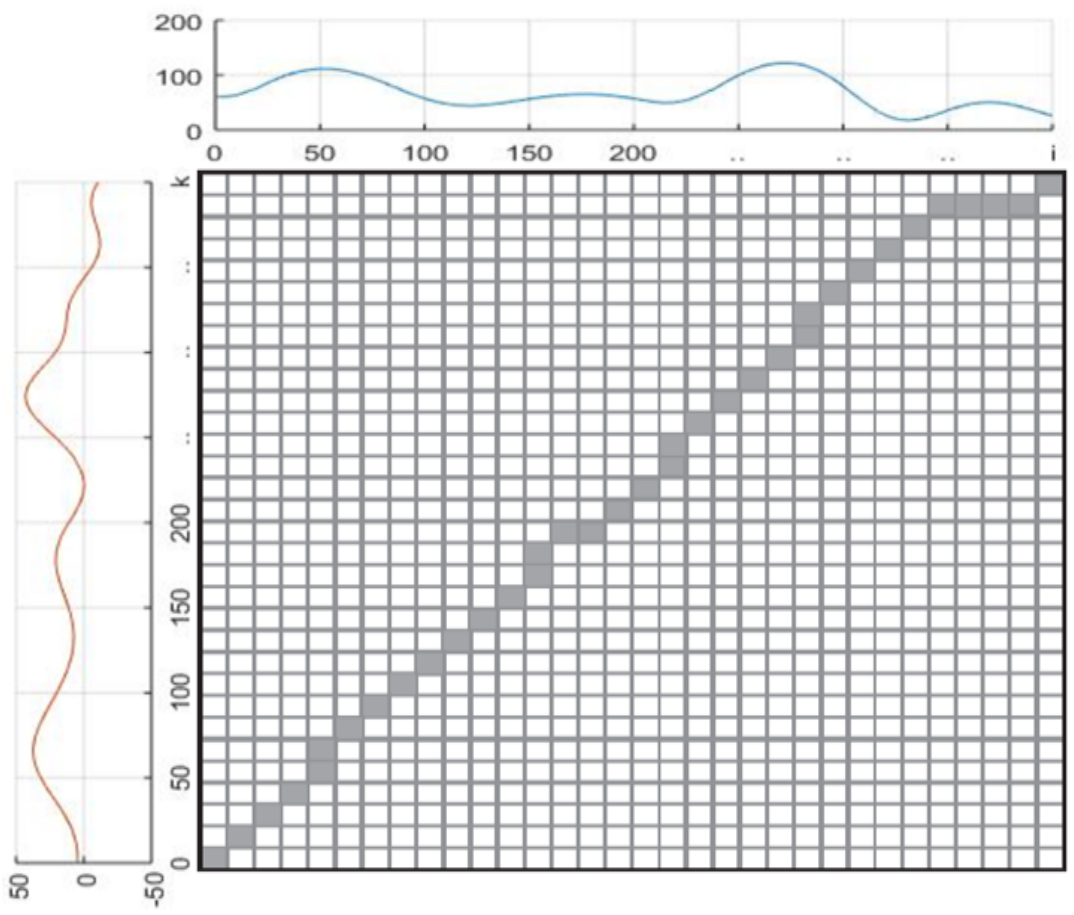

Fig. 1. An optimal warping path. (This is a capture of Fig. 3 (B) from [14]).

The total distance $d_w(\delta_p, \delta_q)$ of a warping path $w$ is defined as:

$$d_w(\delta_p,\delta_q) = \sum_{l=1}^{L} d(\delta_{pm_l},\delta_{qn_l}) \tag{2}$$

The DTW distance between two trajectories $\delta_p$ and $\delta_q$ is defined as the minimum total distance among all possible warping paths, which can be found by dynamic programming [14].

$$DTW(\delta_p,\delta_q) = d_{w*}(\delta_p,\delta_q) = \min\{\, d_w(\delta_p,\delta_q)\} \mid w \text{ is a warping path} \tag{3}$$

In this paper, the similarity between rotor angle responses of generators $p$ and $q$ is represented by $DTW(\delta_p, \delta_q)$. This allows a non-linear mapping between two rotor angle curves, even with data loss or communication delays. DTW is highly ranked in pattern recognition and computer vision fields. It has been widely used in time series analysis, (partial) shape matching, speech recognition, and online signature verification [15]. In [16]-[17], DTW is tested against Euclidean distance for small data size and is found to provide smaller out-of-sample error rate as a result of its improved similarity measure.

Given the coherency of generators, the optimal number of coherent groups $k$ is selected by minimizing inter-coherent group distances [23].

### B. Buses Clustering for Controlled Islanding

After clustering generators, the next step is to find an optimal cut set for controlled islanding with generator coherency information as a constraint. The main task is to allocate non-generator buses to coherent generator groups based on certain metric(s). A candidate approach is spectral clustering, which builds on the concept of minimum graph-cut [18].

Power network can be represented as a weighted graph $G=(V, E, W)$ with vertices ($V$) and edges ($E$) resembling buses and branches (lines or transformers), respectively. To replicate characteristics of the power grid, each edge in the graph is assigned a certain weight ($W$), which can be any system parameter depending on the targeted application. In this work, power flows through branches are used as the weighting factors. Further, to accommodate system losses, weights are evaluated by averaging power flows measured at both sides of the lines as follows.

$$W_{ij} = W_{ji} = \begin{cases} avg(|P_{ij}|, |P_{ji}|) & i \neq j \\ 0 & i = j \end{cases} \tag{4}$$

where $P_{ij}$ and $P_{ji}$ are the active power flows measured at terminal $i$ and $j$ of branch $i$-$j$, respectively. The weight matrix in (4) takes into account the dynamic characteristic of power network as power flow changes with system operating conditions. After evaluating the weight matrix, an un-normalized Laplacian matrix, $L$, can be formulated with its element $L_{ij}$ calculated as:

$$L_{ij} = \begin{cases} -W_{ij} & i \neq j \\ d_i = \sum_{j=1}^{n} W_{ij} & i = j \end{cases} \tag{5}$$

where $d_i$ is the sum of weights of all edges connected to node $i$. To make graphs with different weights comparable, the Laplacian matrix can be normalized as $L_N=D^{-1/2}L_{ij}D^{-1/2}$ [19], where $D$ is a diagonal degree matrix with $d_i$ as its diagonal entries.

Given the generator coherent groups, we apply spectral clustering to further cluster buses for controlled islanding. To incorporate generator coherency information as a constraint in spectral clustering, two types of linkages can be introduced: Must Link (ML) and Cannot Link (CL). Must Link constraints ensure the coherent generators remain on the same island while Cannot Link (CL) keeps the non-coherent generators on different islands. A linkage constraint matrix $Q$ is defined as:

$$Q_{ij} = \begin{cases} +1 & i, j \in ML \\ -1 & i, j \in CL \\ 0 & else \end{cases} \tag{6}$$

Let $\boldsymbol{u} \in \{-1, +1\}^N$ be an island indicator vector for $N$ buses; where, $u_i=+1$ if bus $i$ belongs to island + and $u_i=-1$ if bus $i$ belongs to island −. An index $\boldsymbol{u}^T Q \boldsymbol{u} = \Sigma_{ij}\, u_i u_j Q_{ij}$ can be defined to determine how well constraints in $Q$ are satisfied by the assignment $\boldsymbol{u}$. The greater the value of $\boldsymbol{u}^T Q \boldsymbol{u}$, the more satisfied the coherency constraints $Q$ are by the associated indicator vector $\boldsymbol{u}$ [20]. Variables $u_i$ and $Q$ can be relaxed for more than two islands and soft constraints as $\boldsymbol{u} \in R^N$ and $Q \in R^{N\times N}$ respectively. If $Q_{ij}>0$, then buses $i$ and $j$ should be on the same island and if $Q_{ij}<0$ buses $i$ and $j$ should be placed on different islands. Similar to the normalized Laplacian matrix, $L_N$, constraint matrix $Q$ can also be normalized as $Q_N=D^{-1/2}QD^{-1/2}$. Finally, the association of non-generator buses to already identified generator groups can be obtained by solving the following constrained optimization problem [19]:

$$\arg_v \min\ v^T L_N v \tag{7}$$
$$s.t. \qquad v^T Q_N v > \beta, v^T v = vol, v \neq D^{1/2} 1$$

where $\boldsymbol{v}^T L_N \boldsymbol{v}$ is the cost of the spectral cut, $\beta$ is the satisfaction threshold for constraints, and $vol= \Sigma_i^N\, d_{ii}$ is the volume measure of the graph. $\boldsymbol{v}^T\boldsymbol{v}=vol$ is used to normalize $\boldsymbol{v}$ and $\boldsymbol{v} \neq D^{1/2}\boldsymbol{1}$ is used to avoid trivial solutions with $\boldsymbol{1}$ as a constant vector whose entries are 1s. The relaxed island indicator vector $\boldsymbol{u}$ can be recovered from $\boldsymbol{v}$ as $\boldsymbol{u}=D^{-1/2}\boldsymbol{v}$. The optimal solution of (7) can be obtained using Karush-Kuhn-Tucker theorem [21] by solving the following generalized eigenvalue problem:

$$L_N v = \lambda (Q_N - \frac{\beta}{vol} I) v \cdot \tag{8}$$

After normalizing eigenvectors associated with positive eigenvalues using $v \leftarrow \frac{v}{\|v\|}\sqrt{vol}$ and $k$ being the coherent generator groups obtained through the proposed algorithm, $k-1$ eigenvectors with lowest eigenvalues are selected. Finally, the $k$-medoids algorithm [22] can be applied, on a matrix $V^*$ having $k-1$ eigenvectors as columns. It will allocate non-generator buses to $k$ islands.

## III. An Adaptive Controlled Islanding Framework

To form self-sustained islands, generator coherency and generation/load imbalance need to be considered. In this paper, we treat generator coherency as the constraint and power flow disruption as the minimization objective as shown in (9).

$$Min_{S_1, S_2 \subset G} \Big( \sum_{i \in S_1, j \in S_2} \Big( \frac{|P_{ij}| + |P_{ji}|}{2} \Big) \Big) \tag{9}$$

Here, $S_1$ and $S_2$ are any disjoint groups of generators. Formation of islands with this objective function avoids overloading of lines within the island [17].

Cascading outages can initiate electromechanical oscillations in power systems. As shown in Fig. 2, two cascaded outages occurred at t=5sec and t=7sec, and one generator lost synchronizm. The system eventually became unstable at t=11.45 sec. An efficient islanding scheme should separate generators with different behavior and ensure that coherent generators remain on the same island: (1) to improve the transient stability, (2) to reduce the chances of further outages.

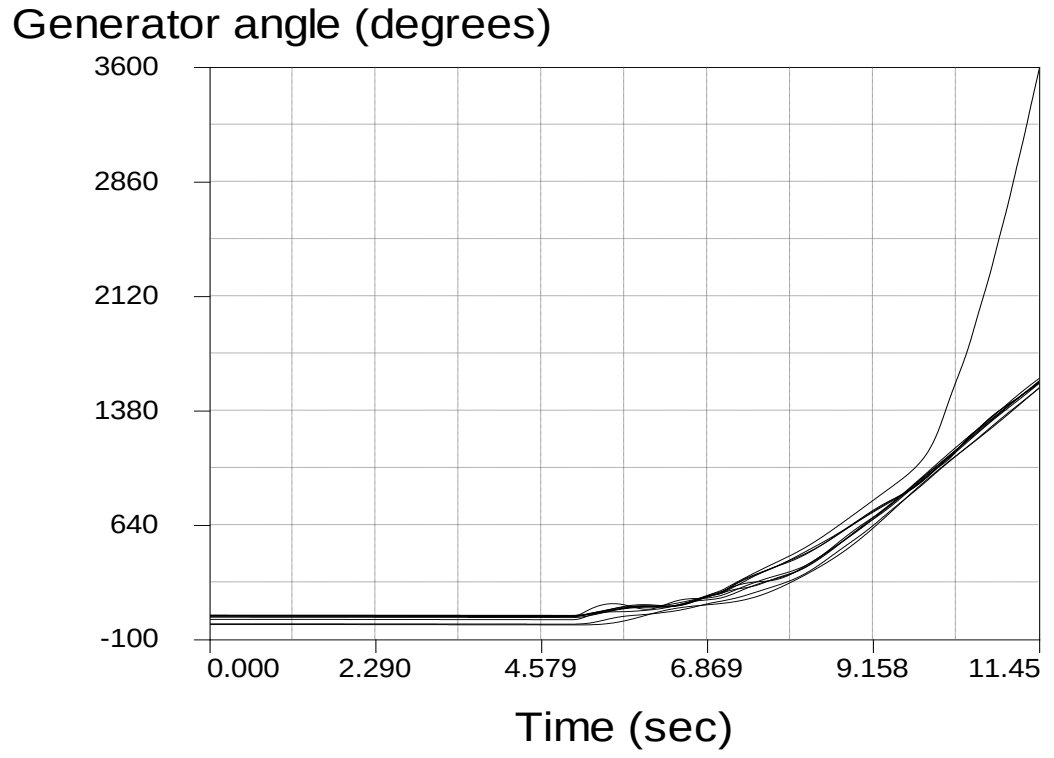

Fig. 2. Generators response following cascaded outages

The proposed adaptive controlled islanding scheme, as shown in Fig. 3, can be implemented using following steps:

*Step 1*: Estimation of generators' rotor angles based on PMU measurements of voltage and current at each generator terminal bus.
*Step 2*: Similarity evaluation between generators rotor angle responses using algorithm proposed in Section II-B. It defines a matrix of similarity index for each pair of generators.
*Step 3*: Optimal number of coherent groups ($k$) selection by minimizing inter-coherent group distances [23]. It provides the number of unique coherent groups.
*Step 4:* Grouping of generators using $k$-means into $k$ coherent groups, obtained from *step 3*, and building a coherency constraint matrix $Q$ using (6).
*Step 5*: Formation of graph $G=(V, E, W)$ using power flow results.
*Step 6*: Building edges' weight matrix $W$ and Laplacian matrix $L$ using (4) and (5) respectively.
*Step 7*: Solving constrained optimization problem in (7) by finding eigenvalues in (8).
*Step 8*: Ignore eigenvectors associated with non-positive eigenvalues. After normalizing the remaining eigenvectors, only consider those eigenvectors that are associated with smallest $k$-$1$ eigenvalues.
*Step 9*: Allocation of non-generator buses to generator groups using $k$-medoids algorithm on the matrix consists of $k-1$ eigenvectors. The opening of all circuit breakers installed on lines whose terminal buses are in distinct groups will eventually form the desired islands.

## IV. Simulation Results

The proposed methodology was tested through dynamic simulations of IEEE 39 bus system. Cascading outages were created using TSAT tools. Time domain simulations show how the proposed methodology can help in minimizing the impact of cascading outages and avoiding blackouts. Furthermore, load/generation imbalance comparison demonstrates the superior performance of proposed methodology comparing to benchmarks.

### A. *Case 1: Comparison with Correlation based Method*

In this case, a 3-phase fault was applied on line 17-16 near bus 17 at t=5sec and cleared after 150ms with the tripping of the corresponding line. Another line 2-1 was tripped at t=7sec following a 3-phase fault of 280ms duration [23]. These cascading outages eventually led the system to lose synchronism at t=12.36 sec as shown in Fig. 4(a). Voltage magnitudes at buses also went very low resulting in a blackout as can be seen in Fig. 4(b).

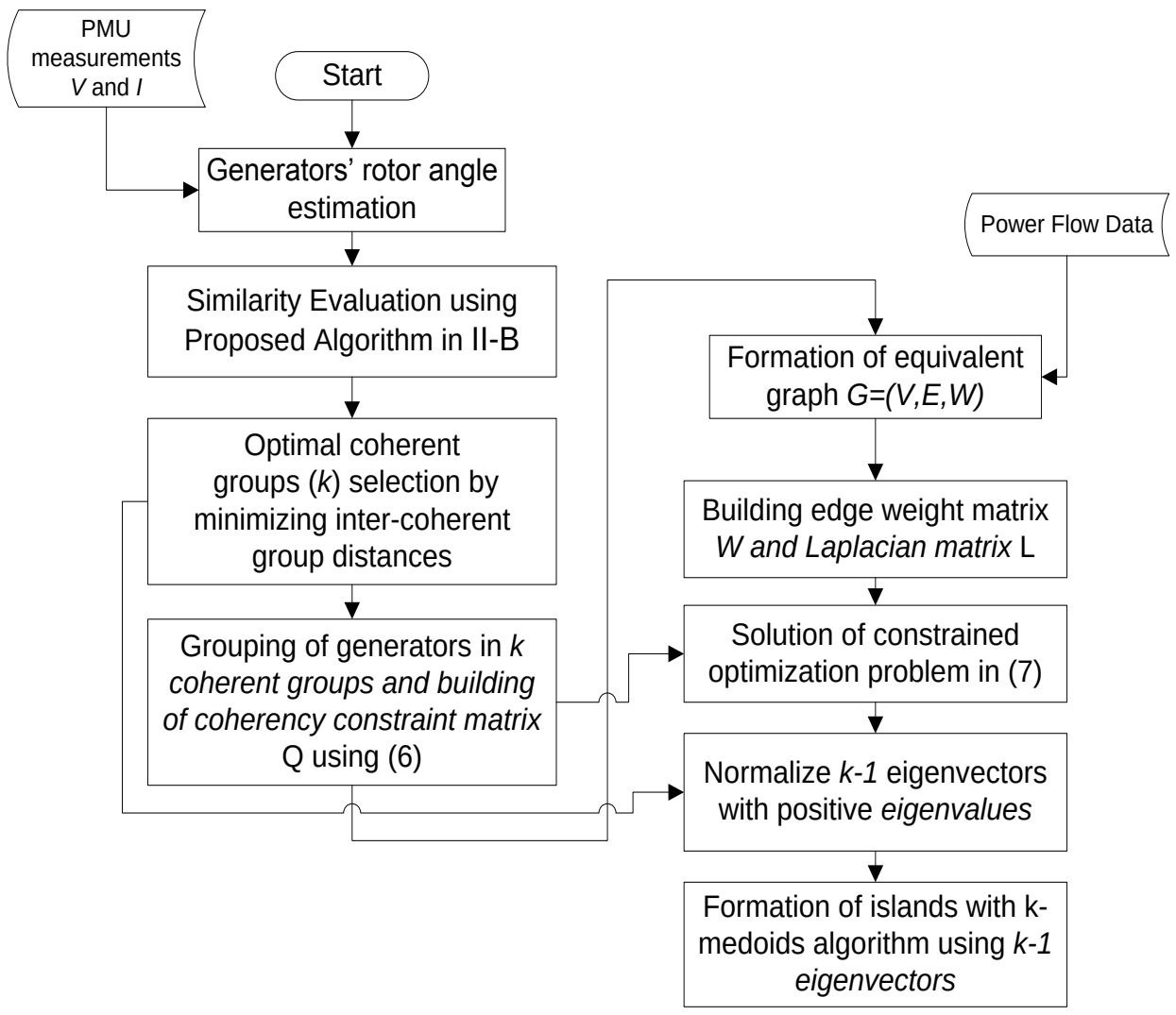

Fig. 3. Algorithm 1: adaptive controlled islanding

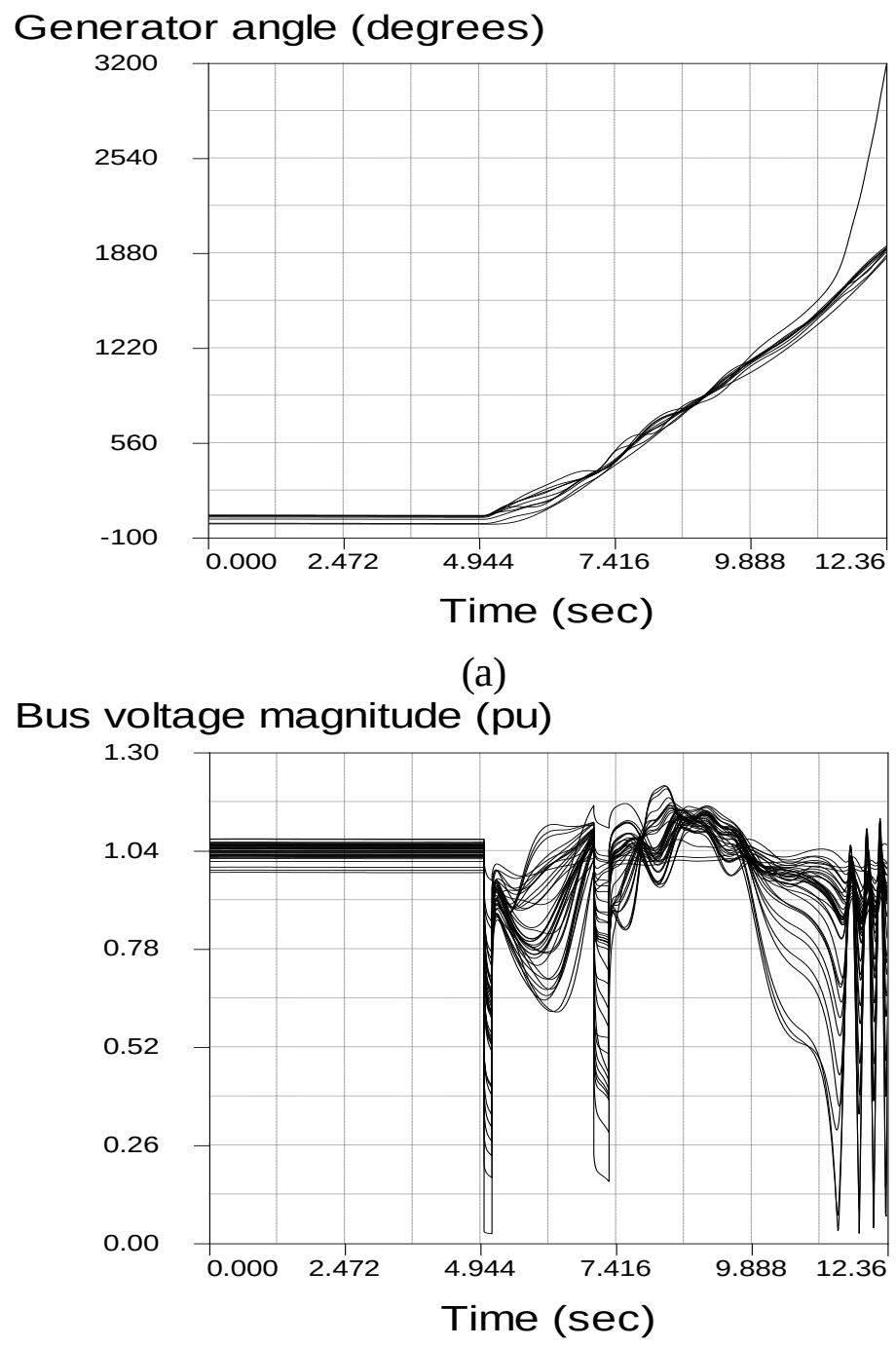

(a)

(b)

Fig. 4. System losing synchronism and becoming unstable

The loss of synchronism and voltage violations are clear indications that the system should be split. In a practical implementation, the timing of splitting is determined by the system operator. Moreover, it depends on the vulnerability analysis performed after severe disturbances [19]. In this paper, we implemented intentional islanding at t=9sec following two cascading outages. The proposed approach provides a suitable islanding solution using online coherency and pre-fault power flow conditions. The proposed generators coherency algorithm identified two coherent generator groups as *(G1, G8, G9)* and *(G2, G3, G4, G5, G6, G7)*. We used this information and solved a constrained spectral clustering problem as described in Section III. Table I shows the allocation of non-generator buses to coherent generator groups. It suggests that the breaker on line 3-4 should be opened to form two islands as shown in Fig. 5. 74.76 MW active power is disrupted. Generators rotor angles also show the clear formation of two coherent groups after islanding as shown in Fig. *6*(a). Voltage magnitude at buses is within limits as can be seen in Fig. *6* (b). The numerical results suggest that Algorithm 1 is capable of avoiding system-wide blackouts by keeping voltages at buses within limits and maintaining generators synchronism.

TABLE I.
ALLOCATION OF NON-GENERATOR BUSES

| Island 1 | Island 2 |
|---|---|
| 2,3,17,18,25,26,27,28,29,**30**,**37**,**38** | 1,4,5,6,7,8,9,10,11,12,13,14,15,16,19,20,21,22,23,24,**31**,**32**,**33**,**34**,**35**,**36**,39 |

To check the quality of islanding, active and reactive power generation capacities and load demands were evaluated for each independent island as presented in Table II. Generators in each island are capable of fulfilling local demand after islanding. Hence, the proposed online coherency algorithm is capable to identify suitable generator groups, which can be used as dynamic constraint for intentional islanding at the expense of minimum load shed to avoid blackout.

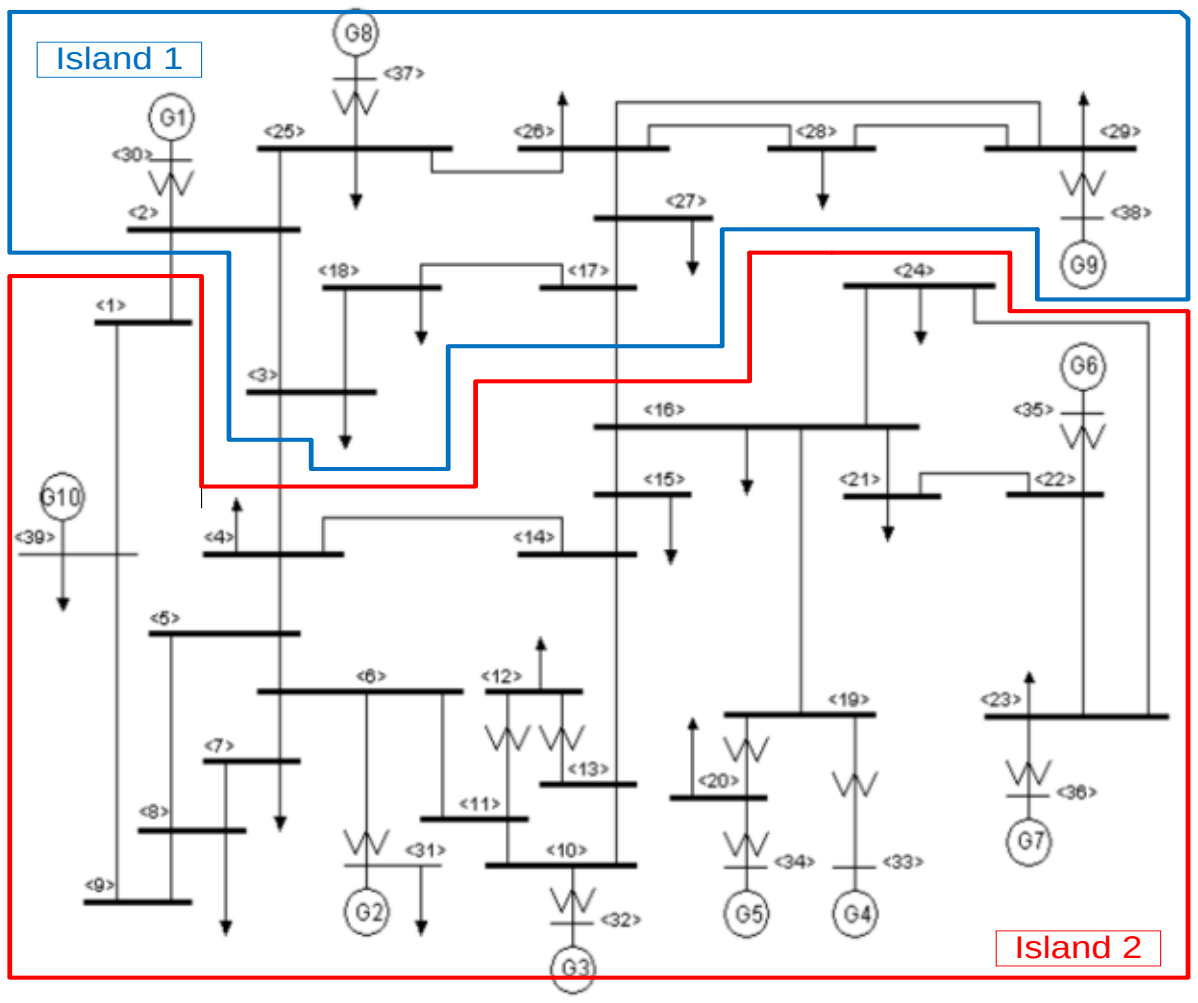

Fig. 5. Formation of two islands

The correlation-based method proposed in [24] was carried out as a benchmark. It calculates the correlation coefficient for each pair of generators and splits them based on the average correlation value.

The correlation-based method identified three coherent groups as *(G2, G3, G10), (G4, G5, G6, G7), (G1, G8, G9)*. The generation capacity of island 1 is below the local demand of the island. About 145.1 MW load needs to be shed as shown in Fig. 7 (b) with the red color area at the top of the load bar.

Moreover, breakers on lines 3-4 and 14-15 should be opened to split the system into three islands. On the other hand, Algorithm 1 sheds no loads with fewer islands and breaker operations. A complete comparison of Algorithm 1 and correlation method based islanding is in Table IV.

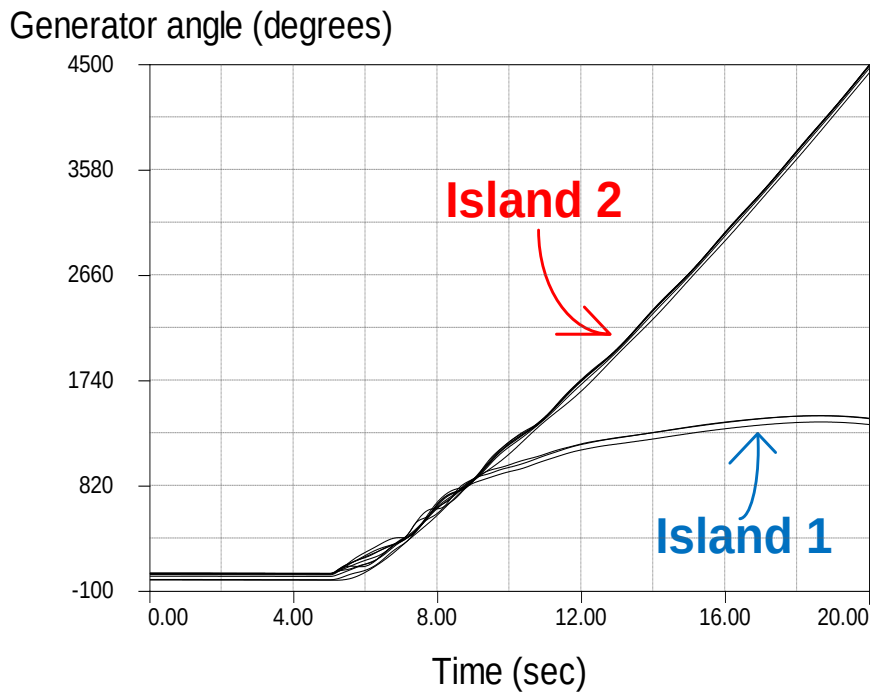

(a)

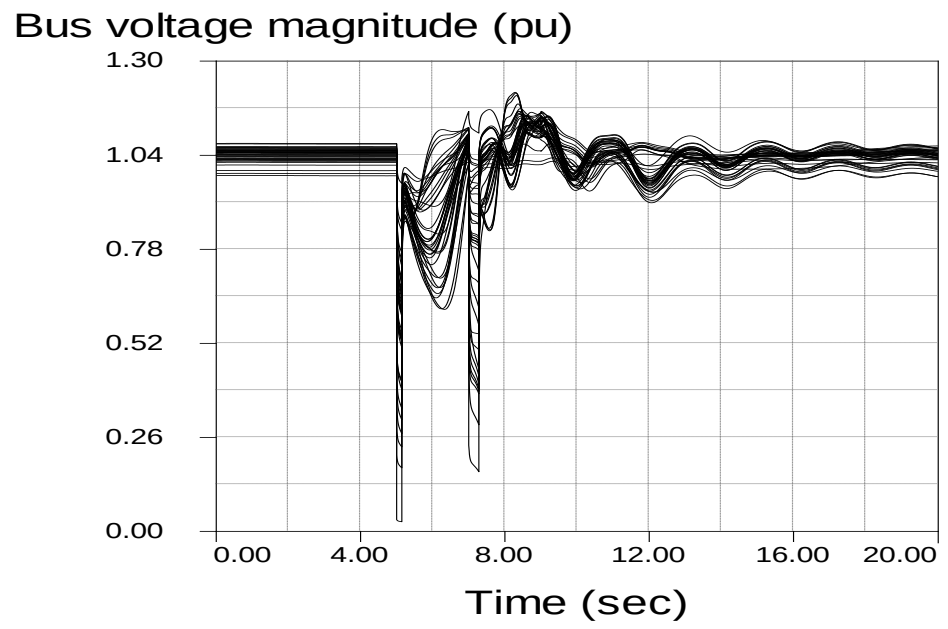

(b)

Fig. 6. Generators rotor angle responses and voltage profiles at system buses after implementing proposed islanding scheme

TABLE II.
ACTIVE AND REACTIVE POWER BALANCES IN EACH ISLAND USING PROPOSED APPROACH

| **Island** | **Active Power Generation Capacity ($P_G$ in p.u)** | **Active Power Load Demand ($P_L$ in p.u)** | **Reactive Power Generation Capacity ($Q_G$ in p.u)** | **Reactive Power Load Demand ($Q_G$ in p.u)** |
|---|---|---|---|---|
| **1** | 16.20 | 16.13 | +24 to -15 | 3.266 |
| **2** | 45.73 | 45.36 | +59 to -38 | 14.73 |

TABLE III.
ACTIVE AND REACTIVE POWER BALANCES IN EACH ISLAND USING CORRELATION BASED ALGORITHM

| **Island** | **Active Power Generation Capacity ($P_G$ in p.u)** | **Active Power Load Demand ($P_L$ in p.u)** | **Reactive Power Generation Capacity ($Q_G$ in p.u)** | **Reactive Power Load Demand ($Q_G$ in p.u)** |
|---|---|---|---|---|
| **1** | 22.239 | 23.69 | +31 to -20 | 7.866 |
| **2** | 23.50 | 21.595 | +28 to -18 | 6.858 |
| **3** | 16.20 | 16.13 | +24 to -15 | 3.266 |

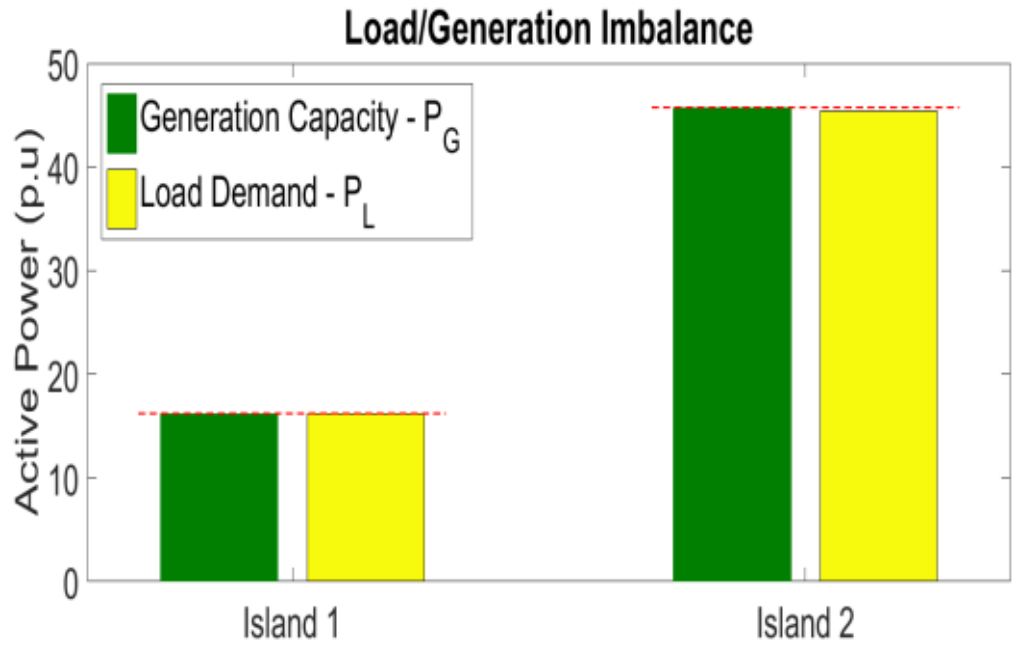

(a)

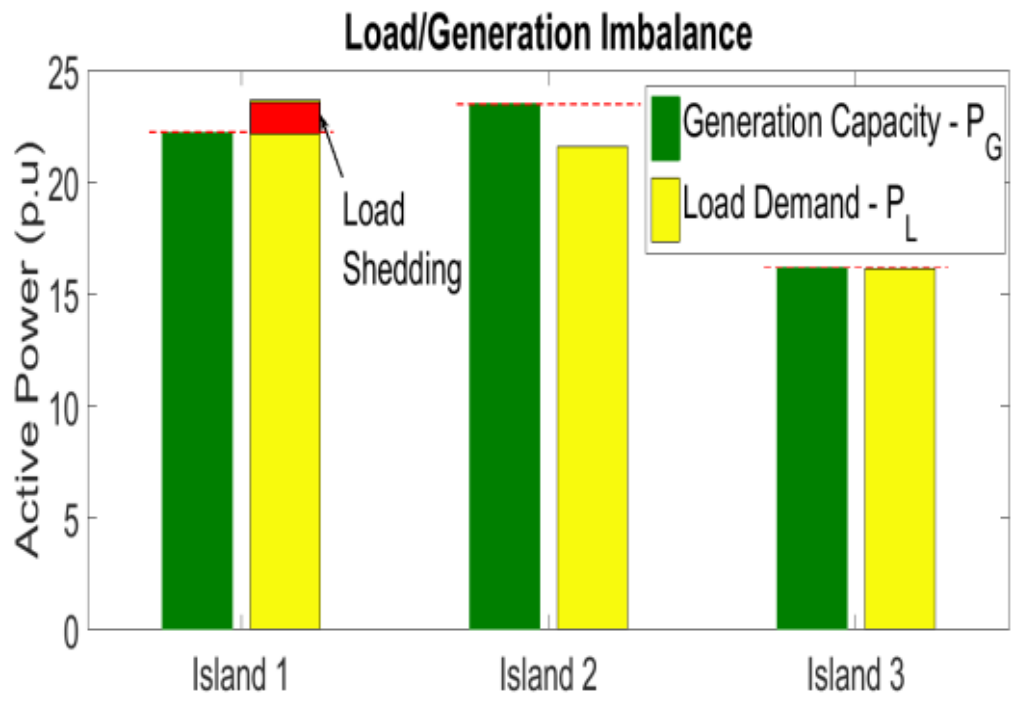

(b)

Fig. 7. Active power load shedding comparison; (a) Proposed method, (b) Correlation based method

TABLE IV.
PERFORMANCE COMPARISON BETWEEN THE PROPOSED AND CORRELATION BASED ALGORITHMS

| **Method** | **Proposed Algorithm based Islanding** | **Correlation Coefficient based Islanding** |
|---|---|---|
| **No. of Lines Cut** | 1 | 2 |
| **No. of Island Formed** | 2 | 3 |
| **Load/Generation Imbalance** | 44.27 MW | 342.6 MW |
| **Load Shed** | 0 MW | 145.1 MW |

### *B. Case 2: Comparison with Community Detection Method*

A 3-phase fault was simulated on line 13-14 near bus 13 at t=5sec and cleared after 150ms with the tripping of the line. Another 3-phase fault of 6 cycles duration was simulated in the middle of the line 16-17 at t=7sec [2]. Following these cascading outages, the system loses synchronism at t=11.45 sec, and voltage magnitudes also go beyond permissible limits as shown in Fig. 8(a) and Fig. 8(b) respectively.

The proposed coherency algorithm identified two generators groups as *(G1, G2, G3, G8, G9)* and *(G4, G5, G6, G7)*. Solving the constrained spectral clustering problem, we got the allocation of non-generator buses as presented in Table V. According to the allocation, the breaker on line 14-15 should be opened to split the system into two islands as shown in Fig. 9. 33.41 MW power was disrupted. Rotor angle trajectories shown in Fig. 11(a) indicate the synchronism of generators after islanding. Voltage magnitudes are also within limits as shown in Fig. 11(b).

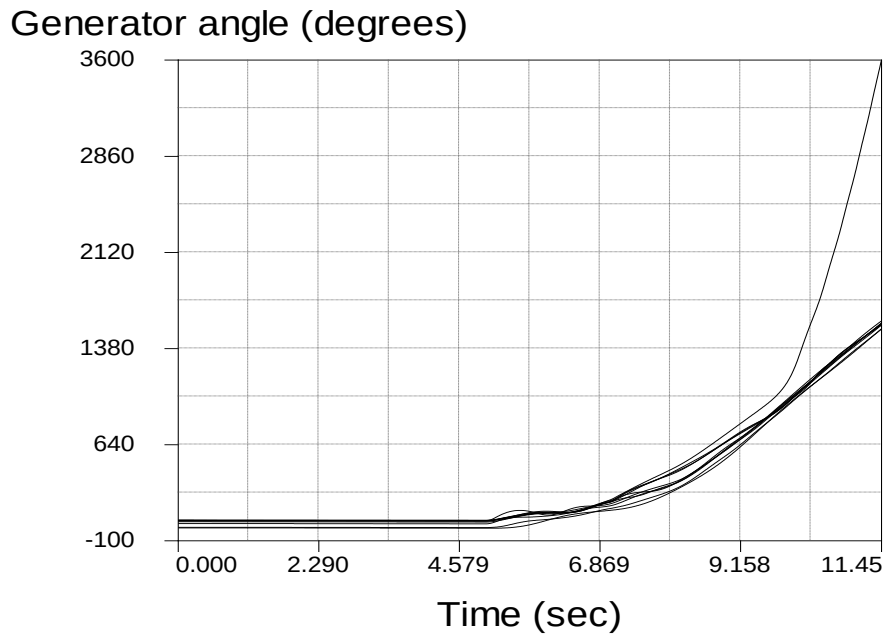

(a)

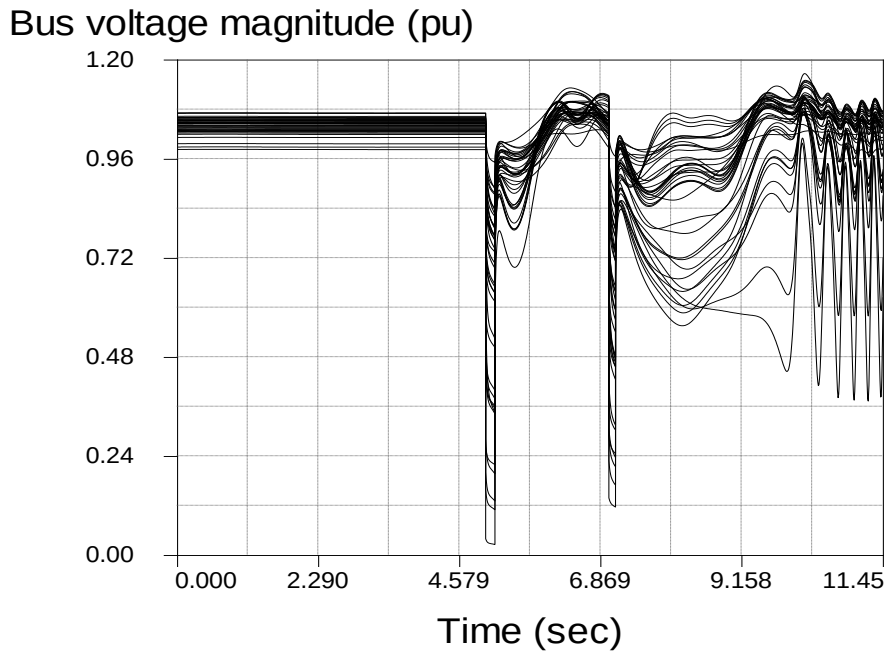

(b)

Fig. 8. System losing synchronism and becoming unstable

TABLE V.
ALLOCATION OF NON-GENERATOR BUSES

| Island 1 | Island 2 |
|---|---|
| 15,16,19,20,21,22,23,24,**33**,**34**,**35**,**36** | 1,2,3,4,5,6,7,8,9,10,11,12,13,14,17,18,25,26,27,28,29,**30**,**31**,**32**,**37**,**38**,39 |

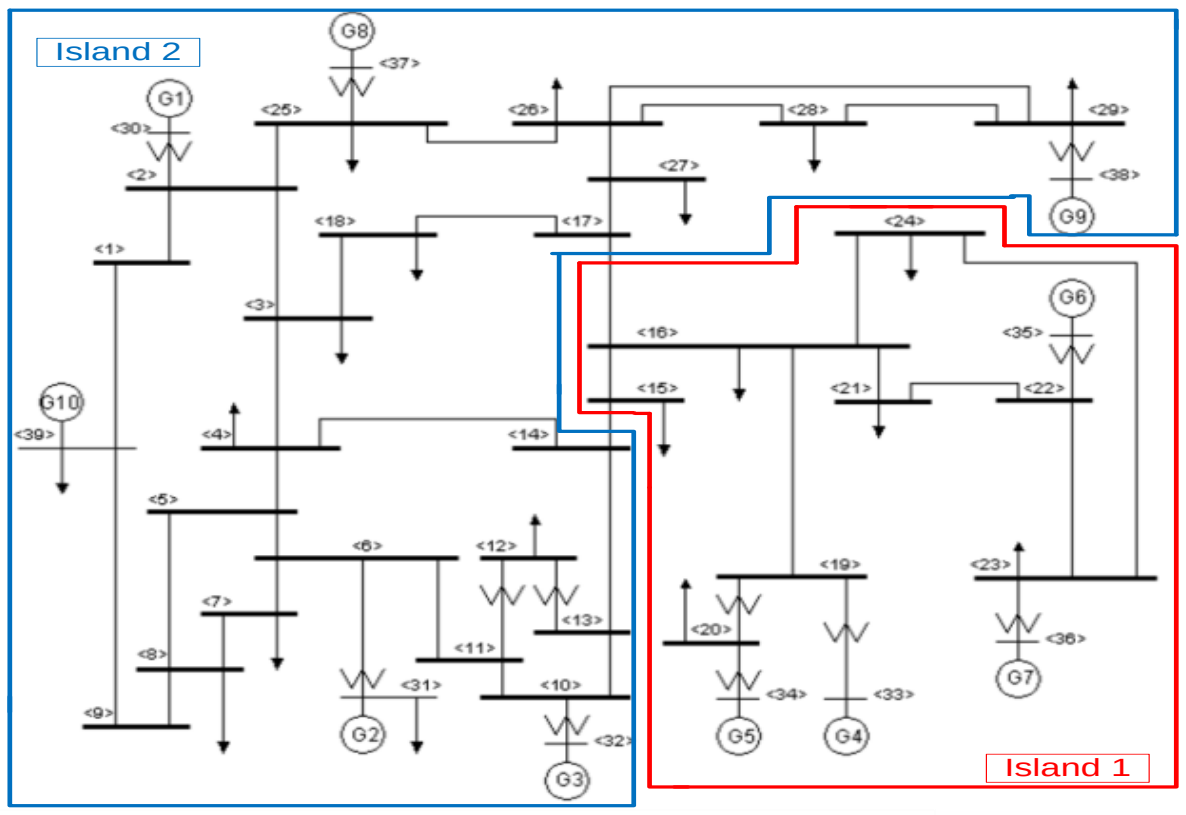

Fig. 9. Formation of two islands

Active and reactive power generation/load imbalance was evaluated for each island as shown in Table VI. Generators in island 1 were capable of fulfilling the load demand. However, 137.7 MW load needed to be shed in island 2 for stable and balanced operation as shown in Fig. 10(a).

We also carried out community detection method introduced in [2], and results are summarized in Table VII. Community detection method identified three coherent generator groups as *(G2, G3)*, *(G4, G5, G6, G7)*, *(G1, G8, G9, G10)*. The active power generation capacities of island 1 and island 3 are less than the demand of each island. Consequently, 50.6 MW and 96.43 MW loads are shed in island 1 and 3 respectively as shown in Fig. 10(b). Breakers on lines 3-4, 8-9 and 14-15 should be opened to split the system into three islands. A complete comparison of Algorithm 1 and community detection method based islanding can be seen in TABLE VIII, which also indicates superior performance of proposed algorithm.

## V. PERFORMANCE OF PROPOSED APPROACH WITH PARTIAL OBSERVABILITY

The performance of online PMU measurements based algorithms is sensitive to partial loss or delay. In PMU based WAMS, communication link failure is common, which may lead the system to be partially observable. Monitoring and control with incomplete information may result in misoperation. Hence, it is important to ensure that the coherency identification method is robust to some extent against partial loss/delay of PMU data. Moreover, due to the ever-decreasing cost of PMUs, as compare to benefits gained in the form of increased system observability, their deployment is massively increasing. This increased dependency on PMUs also poses some challenges for online approaches in case of partial observability of the system. This area has not been widely explored, specifically for online coherency identification application. Some researchers also reported it as the limitation of their proposed coherency identification approach [2].

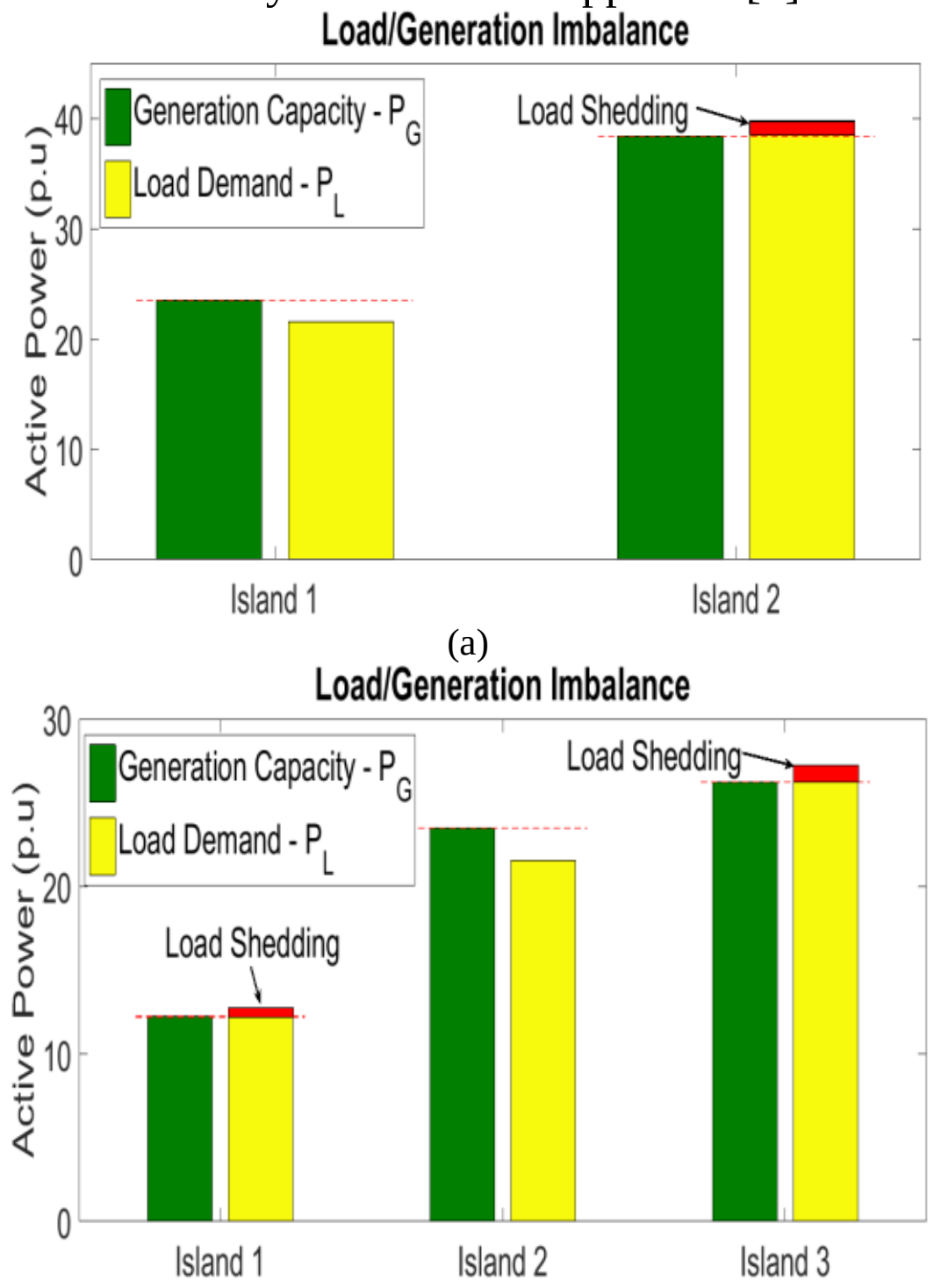

Fig. 10. Active power load shedding comparison; (a) Proposed method, (b) Correlation based method

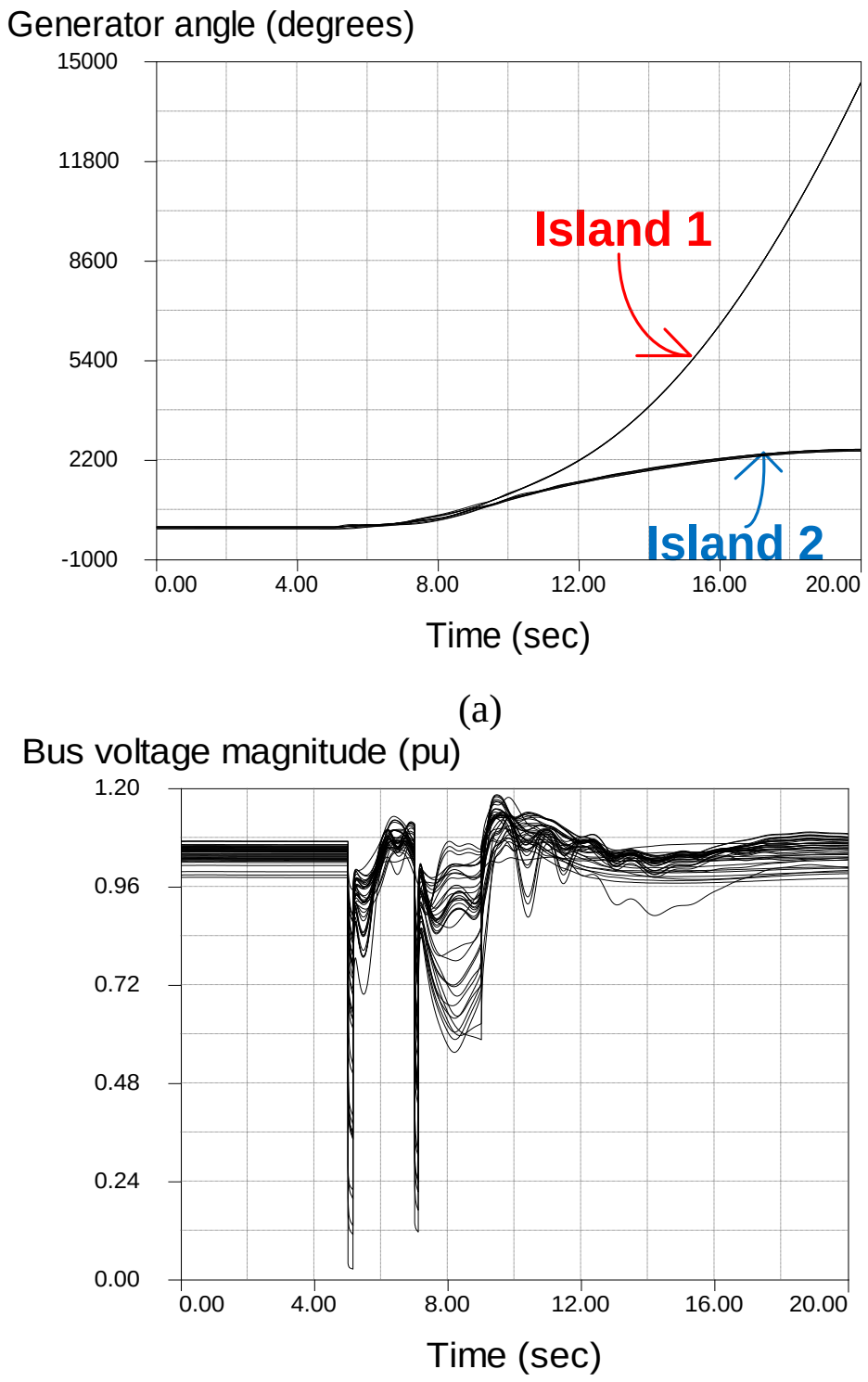

(a)

(b)

Fig. 11. Generators rotor angle responses and voltage profiles at system buses after proposed islanding

TABLE VI.
ACTIVE AND REACTIVE POWER BALANCES IN EACH ISLAND USING PROPOSED APPROACH

| Island | Active Power Generation Capacity ($P_G$ in p.u) | Active Power Load Demand ($P_L$ in p.u) | Reactive Power Generation Capacity ($Q_G$ in p.u) | Reactive Power Load Demand ($Q_G$ in p.u) |
|---|---|---|---|---|
| **1** | 23.50 | 21.59 | +28 to -18 | 7.16 |
| **2** | 38.43 | 39.81 | +55 to -35 | 10.83 |

TABLE VII.
ACTIVE AND REACTIVE POWER BALANCES IN EACH ISLAND USING COMMUNITY DETECTION BASED ALGORITHM

| Island | Active Power Generation Capacity ($P_G$ in p.u) | Active Power Load Demand ($P_L$ in p.u) | Reactive Power Generation Capacity ($Q_G$ in p.u) | Reactive Power Load Demand ($Q_G$ in p.u) |
|---|---|---|---|---|
| **1** | 12.229 | 12.735 | +16 to -10 | 5.366 |
| **2** | 23.50 | 21.58 | +28 to -18 | 6.864 |
| **3** | 26.21 | 27.175 | +39 to -25 | 5.766 |

TABLE VIII.
PERFORMANCE COMPARISON BETWEEN THE PROPOSED AND COMMUNITY DETECTION BASED ALGORITHMS

| Method | Proposed Algorithm based Islanding | Community Detection based Islanding |
|---|---|---|
| **No. of Lines Cut** | 1 | 3 |
| **No. of Island Formed** | 2 | 3 |
| **Load/Generation Imbalance** | 328.2 MW | 339.03 MW |
| **Load Shed** | 137.7 MW | 147.03 MW |

The proposed online coherency approach is applicable in the case of partial observability of the system due to its non-linear nature of similarity computation as explained in Section II-A. Consider case 1 mentioned in Section IV, where we have cascaded outages of line 17-16 and 2-1 at t=5sec and t=7 sec respectively as shown in Fig. 2. We considered PMUs on generator buses only. To analyze the performance of proposed online coherency algorithm for a partially observable system, we intentionally removed the initial measurement points for each PMU. Fig. 12 shows the experimental results. The green color in each curve indicates the lost part of PMU data. We determined the coherency through proposed online approach. After determining the coherency with such incomplete PMU data, we compared the coherency results with the results obtained without any loss of measurements. The algorithm allows accommodating the partial loss of data to some extent and still gives us the same coherency results. Thus, the proposed online coherency approach is robust to a considerable extent for loss/delay of PMU data.

## VI. CONCLUSION

Splitting a power system into self-sustained islands is the last resort to maintain transient stability. This paper presents a novel methodology for generator coherency identification. It uses post-fault rotor angle trajectories of generators for coherency determination. For non-generator buses allocation, constrained spectral clustering is applied to minimize power flow disruption, considering generator coherency as a constraint. Future work includes 1) allocation of non-generator buses based on multiple constraints like restoration constraint, thermal limits of transmission lines, etc., in addition to generator coherence constraint; 2) prevention of blackouts using energy storage system without going into islanding operation mode.

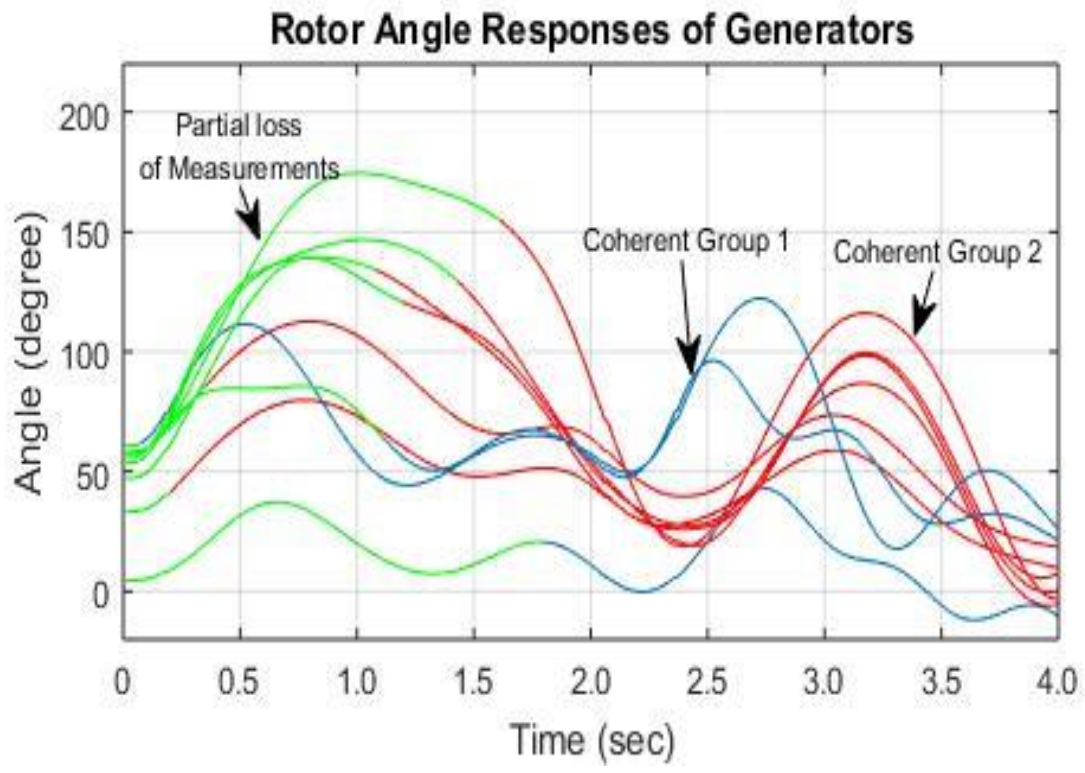

Fig. 12. Performance of proposed coherency algorithm with partial loss of PMU data